\documentclass{article}
\usepackage{amsmath,graphicx}
\usepackage{tipa}
\usepackage{tikz}
\usepackage{environ}
\usepackage{textcomp}
\usepackage{microtype}
\usepackage{booktabs}
\usepackage[preprint]{spconf}
\usetikzlibrary{calc} 
\usetikzlibrary{shapes.geometric, arrows}
\tikzstyle{process} = [rectangle, minimum width=1.25cm, minimum height=0.6cm, outer sep=0,text centered, draw=black, fill=white]
\tikzstyle{c} = [circle, minimum size=1.5em,inner sep=0.2,text centered, draw=black, fill=white]
\tikzstyle{cred} = [circle, minimum size=1.5em,inner sep=0.2,text centered, draw=red, fill=white]
\tikzstyle{arrow} = [thick,->,>=stealth]
\tikzstyle{tria} = [regular polygon, regular polygon sides=3,draw, fill=white, text width=1em,inner sep=0.3mm, outer sep=0mm,shape border rotate=-90]
\tikzstyle{tria_r} = [regular polygon, regular polygon sides=3,draw, fill=white, text width=1em,inner sep=0.3mm, outer sep=0mm,shape border rotate=90]
\tikzstyle{io} = [trapezium, trapezium left angle=70, trapezium right angle=70, minimum width=2cm, minimum height=0.1cm, text centered,text width=2cm, draw=black]
\usepackage{lipsum,adjustbox}
\usepackage{standalone}
\usepackage{subcaption}
\usepackage{balance}

\title{Unsupervised Speech Enhancement with speech recognition embedding and disentanglement losses}
\name{Viet Anh Trinh${^1}$  and Sebastian Braun ${^2}$}
\address{
  ${^1}$ The Graduate Center, CUNY, NY, USA\\
  ${^2}$ Microsoft Research, Redmond, WA, USA \\
  vtrinh@gradcenter.cuny.edu, sebastian.braun@microsoft.com}
\begin{document}
\copyrightnotice{ \begin{tabular}[t]{@{}l@{}} To appear in {\it Proc.\ ICASSP 2022, May, 2022, Singapore.} \\
\copyright\ 2022 IEEE. Personal use of this material is permitted. Permission  from IEEE must be obtained for all other uses,  in any current  or  future \\ media,  including reprinting/republishing this material  for advertising or promotional purposes, creating new collective  works, for resale or \\ redistribution to  servers or lists,  or reuse of any copyrighted component of this work in other works. \end{tabular} } 
\ninept
\maketitle
\begin{abstract}
Speech enhancement has recently achieved great success with various deep learning methods. However, most conventional speech enhancement systems are trained with supervised methods that impose two significant challenges. First, a majority of training datasets for speech enhancement systems are synthetic. When mixing clean speech and noisy corpora to create the synthetic datasets, domain mismatches occur between synthetic and real-world recordings of noisy speech or audio. Second, there is a trade-off between increasing speech enhancement performance and degrading speech recognition (ASR) performance. Thus, we propose an unsupervised loss function to tackle those two problems. Our function is developed by extending the MixIT loss function with speech recognition embedding and disentanglement loss. Our results show that the proposed function effectively improves the speech enhancement performance compared to a baseline trained in a supervised way on the noisy VoxCeleb dataset. While fully unsupervised training is unable to exceed the corresponding baseline, with joint super- and unsupervised training, the system is able to achieve similar speech quality and better ASR performance than the best supervised baseline. 
\end{abstract}
\section{Introduction}
\label{sec:intro}

Speech enhancement aims to remove noise and reverberation while maintaining speech and audio quality. 
Although current speech enhancement systems have achieved significant performance with different deep learning architectures, e.g., \cite{weninger2015speech,luo2019conv,reddy21_interspeech,li2021icassp,wang2018supervised, tan2019learning}, there are some challenges arising in this domain. In this study, we focus on two main problems. First, most speech enhancement training datasets are synthetic and created by mixing clean speech and noisy corpora. However, there are domain mismatches between synthetic and real-world recordings of noisy speech or audio. Also, collecting studio-quality training datasets is expensive. Second, speech enhancement systems can remove noise, but they can distort speech signals as well, which in consequence can degrade the performance of subsequent ASR systems. Therefore, this paper examines some previous works and proposes a new loss function to mitigate these two problems.  

There are several unsupervised speech enhancement methods proposed to mitigate the domain mismatch,  e.g., \cite{NEURIPS2020_28538c39,saito2021training,wang2020self,fujimura2021noisy}. However, in speech enhancement, supervised training has not yet been outperformed by unsupervised approaches. MixIT \cite{NEURIPS2020_28538c39} is a special loss function designed for unsupervised speech enhancement. This loss function is a combination of a permutation loss, a reconstruction loss and a speech enhancement loss. The algorithm allows training the speech enhancement system with noisy speech instead of clean speech. MixIT shows its effectiveness in source separation problem but it does not work well for speech enhancement. \cite{saito2021training} aims to improve MixIT by a noise augmentation method, where the training speech sample is made more noisy by adding additional artificial noise. They argued that in the MixIT paper, the distribution of recording noise inside the speech was different from the additional noise that led to the ineffectiveness of MixIT in the speech enhancement task. The authors expected to improve the learning of MixIT by adding a pre-noise from the same distribution as the training noise. 

In this study, we develop a weighted loss function as a solution to the two problems mentioned above. Our function combines speech enhancement, disentanglement and speech recognition (ASR) embedding losses. Taken together, we make three considerable contributions. First, we propose a new loss function that enables using abundant real-world noisy corpora to improve speech enhancement performance. Second, we address the trade-off between speech enhancement and ASR by integrating an ASR embedding loss and a disentanglement term to help separating the outputs of MixIt. Third, we introduce a joint way of leveraging clean and noisy speech combining supervised and unsupervised losses.

\begin{figure}
\centering
\begin{adjustbox}{width=0.8\columnwidth}
\begin{tikzpicture}[every node/.style={font=\footnotesize}]
\node (add1) [c] {$+$};
\node (S)[inner sep=0pt,above of = add1,yshift=-0.3cm] {$S$};
\node (M)[right of = add1,xshift=-0.4cm,yshift=0.2cm]  
    {$M$};
\node (dot1) [c,right of = add1, xshift = -0.1 cm, fill = black,minimum size=0.5em]{};
\node (N)[below of = add1,yshift=0.2cm]  
    {$N$};
\node (encoder) [process, right of =add1, xshift= 1 cm,fill=blue!10] {Encoder};
\node (bottleneck) [process, right of =encoder, xshift= 0.7 cm,fill=yellow!20] {Bottleneck};
\node (decoder) [process, right of =bottleneck, xshift= 0.8 cm,fill=green!30] {Decoder};
\node (G)[right of = decoder,xshift=0.4cm]
    {G};
\node (mul1) [c,right of = G,xshift=-0.2cm] {$*$};
\node (hatS)[right of = mul1,xshift=0.15cm]  
    {$\hat{S}$};
\draw [arrow] (S) -- (add1);
\draw [arrow] (N) -- (add1);
\draw [arrow] (add1) -- (encoder);
\draw [arrow] (encoder) -- (bottleneck);
\draw [arrow] (bottleneck) -- (decoder);
\draw [arrow] (decoder) -- (G);
\draw [arrow] (G) -- (mul1);
\draw [arrow] (mul1) -- (hatS);
\draw (dot1) -- ($(dot1)+(0,0.8)$) ; 
\draw [arrow] ($(dot1)+(0,0.8)$)-| (mul1) ; 
\end{tikzpicture}
\end{adjustbox}
\caption{Supervised network architecture}
\label{fig:supervisedNetwork}
\end{figure}
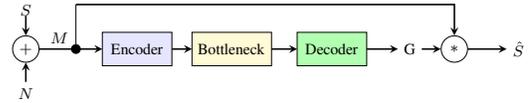

\section{Method}
In this section, we describe the conventional supervised speech enhancement approach used in the baseline and our proposed unsupervised and semi-supervised methods. 
\subsection{Supervised speech enhancement}
\subsubsection{Formulation}
\begin{figure}
\centering
\begin{adjustbox}{width=0.82\columnwidth}
\begin{tikzpicture}[every node/.style={font=\footnotesize}]
\node (add1) [c] {$+$};
\node (X)[inner sep=0pt,above of = add1,yshift=2.6cm] {$X=S + N_{recording}$};
\node (Y)[right of = add1,xshift=-0.4cm,yshift=0.2cm]  
    {$Y$};
\node (N)[below of = add1,yshift=-1.6cm]  
    {$N$};
\node (neuralnetwork) [process, minimum height=4cm, right of =add1, xshift= 1 cm] {Neural network};

\node (N1)[right of = neuralnetwork,xshift=3.5cm]  
    {$N_1$};
\node (S)[ above of = N1, yshift= 0.8 cm]  
    {$\hat{S}$};
\node (N2)[below of = N1,yshift= -0.8 cm]  
    {$N_2$};
\node (w2vS) [process, right of =S, xshift= 0.45 cm] {W2V};
\node (w2vN1) [process, right of =N1, xshift= 0.45 cm] {W2V};
\node (w2vN2) [process, right of =N2, xshift= 0.45 cm] {W2V};
\node (w2vX) [process, above of =w2vS, yshift= 0.8 cm] {W2V};
\node (Semb)[right of = w2vS,xshift=1.2cm]  
    {$\hat{S}_{emb}$};
\node (N1emb)[right of = w2vN1,xshift=1.2cm]  
    {$N_{1-emb}$};
\node (N2emb)[right of = w2vN2,xshift=1.2cm]  
    {$N_{2-emb}$};
\node (Xemb)[right of = w2vX,xshift=1.2cm]  
    {$X_{emb}$};
\node (disenSN1)[above of = N1emb,xshift=0.2,yshift=-0.05cm] {$<\hat{S}_{emb},N_{1-emb}>$} ;
\node (LMSEXS)[above of = Semb,xshift=0.1,yshift=-0.05cm]  
    {$\mathcal{L}_{MSE}(\hat{S}_{emb},X_{emb})$};
\node (LSEN2N)[right of = N,xshift=1cm]  
    {$\mathcal{L}_{SE}(N_2,N)$};
\node (add2) [cred, above of = N1,yshift=-0.05cm] {$\textcolor{red}{+}$};
\node (LSESN1Y)[left of = add2,xshift=-1cm]  
    {$\mathcal{L}_{SE}(\hat{S}+N_1,X)$};
\draw [arrow] (X) -- (add1);
\draw [arrow] (N) -- (add1);
\draw [arrow] (add1) -- (neuralnetwork);
\draw [arrow] ($(neuralnetwork)+(1.0,1.8)$) -- (S); 
\draw [arrow] ($(neuralnetwork)+(1.0,-1.8)$) -- (N2);
\draw [arrow] (neuralnetwork) -- (N1);
\draw [arrow] (S) -- (w2vS);
\draw [arrow] (N1) -- (w2vN1);
\draw [arrow] (N2) -- (w2vN2);
\draw [arrow] (w2vS) -- (Semb);
\draw [arrow] (w2vN1) -- (N1emb);
\draw [arrow] (w2vN2) -- (N2emb);
\draw [dashed,red] (Semb) -- (disenSN1);
\draw [dashed,red] (N1emb) -- (disenSN1);
\draw [dashed,red] (LMSEXS) -- (Semb);
\draw [dashed,red] (LMSEXS) -- (Xemb);
\draw [arrow] (w2vX) -- (Xemb);
\draw [arrow] (X) -- (w2vX);
\draw [dashed,red,rounded corners=5pt] (LSEN2N) -- (N);
\draw [dashed,red,rounded corners=5pt] (LSEN2N) -| (N2);
\draw [dashed,red,rounded corners=5pt] (LSESN1Y) -| (X);
\draw [dashed,red] (S) -- (add2);
\draw [dashed,red] (N1) -- (add2);
\draw [dashed,red] (LSESN1Y) -- (add2);
\end{tikzpicture}
\end{adjustbox}
\caption{Proposed unsupervised training scheme. The network receives a mixture of a noisy recording $X$ and an additional noise $N$. The three outputs of the network are clean speech $S$ and two noises $N_1$ and $N_2$, respectively. Embeddings of these three signals are extracted using Wav2vec 2.0 (W2V). Red arrows illustrate loss function components}
\label{fig:unsup}
\end{figure}
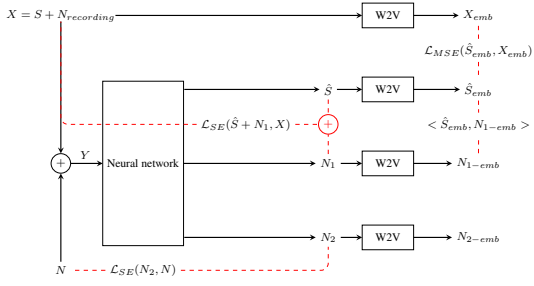

As most speech enhancement systems are supervised, we deployed supervised networks as baselines to have a fair comparison for our proposed unsupervised method.
The supervised architecture is depicted in Figure \ref{fig:supervisedNetwork}. Let $Y$ be the the short-time Fourier transform (STFT) of the input signal, where $k,n$ are the frequency and time indices. The input features to the network are complex compressed spectra given by $|Y|^c \frac{Y}{|Y|}$ with $c\!=\!0.3$. The compressed spectra are fed as real and imaginary in two input channels of the network encoder. 
The output of the network is a complex filter (mask) $G(k,n)$ , and the output signal spectrum $\hat{S}(k,n)$ is obtained by complex multiplication of the predicted filter $G(k,n)$ with the input spectrum $Y(k,n)$. The network is trained by minimizing the supervised signal loss between the target speech $S(k,n)$ and the predicted clean speech $\hat{S}(k,n)$. We use the complex compressed spectral loss \cite{braun2021consolidated}
\begin{align}
  \mathcal{L}_{SE}(S,\hat{S}) = &(1-\lambda)\sum_{k,n}\left||S|^c -|\hat{S}|^c\right|^2 \nonumber \\ &+ \lambda \sum_{k,n}\left||S|^c \frac{S}{|S|} -|\hat{S}|^c \frac{\hat{S}}{|\hat{S}|}\right|^2,
 \label{eq:eqsup}
\end{align}
where $\lambda$ is a weighting between the complex loss and magnitude-based loss \cite{braun2021consolidated}.
\subsubsection{Supervised architecture}
We use the Convolutional Recurrent U-net for Speech Enhancement (CRUSE) \cite{braun2021towards} architecture. The enhancement system of CRUSE is illustrated in Fig.~\ref{fig:supervisedNetwork} and the detailed network architecture is shown in Fig.~\ref{fig:detailedNetwork}. The two additional decoder branches in red required for unsupervised MixIT training are explained in the following section. CRUSE has a symmetric encoder-decoder and a gated recurrent unit (GRU) \cite{cho2014properties} as bottleneck. The encoder has $L$ convolutional layers, where the channel number doubles, and the feature size is reduced by half each layer by striding. The decoder has $L$ deconvolutional layers, mirroring the encoder layers by upsampling and decreasing channels. There are skip connections between corresponding layers of encoder and decoder with inserted $1\times1$ convolutions. The bottleneck has $J$ parallel GRU layers. 
\subsection{Unsupervised speech enhancement}
\subsubsection{MixIT loss}
In supervised speech enhancement, the model is trained to make the predicted speech close to the target clean speech. However, in unsupervised speech enhancement, the clean speech is not available. The MixIT loss \cite{NEURIPS2020_28538c39} does not need access to the clean speech target, and is given by 
\begin{align}
    \label{eq:mixit}
  \mathcal{L}_{MixIT-SE} = \min& \bigl[\mathcal{L}_{SE}(\hat{S}+N_1,X) +\mathcal{L}_{SE}(N_2,N), \nonumber \\ &\mathcal{L}_{SE}(\hat{S}+N_2,X) + \mathcal{L}_{SE}(N_1,N) \bigr],
\end{align}
where $X(k,n)$ and $N(k,n)$ are the complex spectra of the 2 inputs: noisy recording and the additional noise. To train using MixIT, the network has to produce three output signal, here the spectra $\hat{S},N_1,N_2$ are the complex spectrogram of the 3 outputs: predicted speech, first noise and second noise, respectively. 
The permutation allows the internal noise inside the real-world noisy recording $X$ and the additional noise either to go to $N_1$ or $N_2$, while the desired speech is expected to always end up in output $\hat{S}$.

\subsubsection{Proposed ASR embedding and disentanglement related loss for unsupervised speech enhancement}

The original MixIT loss was shown to work well in unsupervised speech separation but it did not have a good performance in speech enhancement task \cite{NEURIPS2020_28538c39}. Therefore, we propose two additional constraint terms to the MixIT loss. 

In the first constraint, to prevent degradation of the ASR performance, we use a pre-trained ASR system to extract the embeddings $X_{emb}$ and $\hat{S}_{emb}$ from the noisy speech recording and the enhanced output, and force them to be similar by
\begin{align}
  \mathcal{L}_{emb} = \mathcal{L}_{MSE}(\hat{S}_{emb},X_{emb}).
\end{align}

In the second constraint, we use a disentanglement constraint between speech and noise so that leakage between speech and noise outputs of the decoder is minimized. The ASR embedding of the predicted speech $\hat{S}_{emb}$ is forced to be different from the ASR embeddings of noise 1 $N_{1-emb}$ and noise 2 $N_{2-emb}$ by enforcing orthogonality of the embeddings, i.\,e.\,
\begin{align}
  \mathcal{L}_{dis} = <\hat{S}_{emb},N_{1-emb}>
 + <\hat{S}_{emb},N_{2-emb}>,
\end{align}
where $<>$ specifies the dot product operation. 

Our proposed loss function is the weighted sum of the original MixIT loss and our embedding and disentanglement loss
\begin{align}
  \mathcal{L} = \mathcal{L}_{MixIT-SE} + \alpha_\text{e} \mathcal{L}_{emb} +\alpha_\text{d} \mathcal{L}_{dis}
   \label{eq:equnsup}
\end{align}
where $\alpha_\text{e}$, $\alpha_\text{d}$ are weights for embedding and disentanglement losses, respectively.

\begin{figure}
\centering
\begin{adjustbox}{width=0.82\columnwidth}
\begin{tikzpicture}[every node/.style={font=\footnotesize}]
\node (Y){$Y$};
\node (enc1) [process,minimum width=2.6cm,fill=blue!10,rotate=90,right of = Y,yshift = -1 cm,xshift = -1 cm] {Conv (T,F)};
\node (Cin)[left of = enc1,xshift=0.3cm,yshift=0.4cm]  
    {$C_{in}$};
\node (enc2) [process,minimum width=2.1cm,right of = enc1,xshift=1cm, fill=blue!10,rotate=90] {Conv (T,F)};
\node (C1)[left of = enc2,xshift=0.4cm,yshift=0.4cm]  
    {$C_1$};
\node (enc3) [process,right of = enc2,minimum width=1.6cm,xshift=1.5cm,fill=blue!10,rotate=90] {Conv (T,F)};
\node (C2)[right of = enc2,xshift=-0.4cm,yshift=0.4cm]  
    {$C_2$};
\node (CL1)[right of = enc2,xshift=0.75cm,yshift=0.4cm]  
    {$C_{L-1}$};
\node (CL)[right of = enc3,xshift=-0.5cm,yshift=1.1cm]  
    {$C_L$};
\node (text1)[right of = enc1,xshift=-1.1cm,yshift=1.7cm]  
    {Downsampling
    striding};
\node (flatten1) [io,minimum width=2.6cm,right of = enc3,xshift=0.7cm,rotate=90,fill=yellow!20] {$flatten$};
\node (gru1) [process,minimum width=0.87cm,right of = flatten1,xshift=-0.45cm,rotate=90,fill=yellow!20] {...J};
\node (gru2) [process,minimum width=0.87cm,above of = gru1,yshift=-0.12cm,rotate=90,fill=yellow!20] {GRU};
\node (gru3) [process,minimum width=0.87cm,below of = gru1,yshift=0.12cm,rotate=90,fill=yellow!20] {GRU};
\node (reshape) [io,minimum width=2.6cm,minimum height=0.2cm,right of = gru1,xshift=-0.45cm,rotate=90,trapezium left angle=110, trapezium right angle=110,fill=yellow!20]{reshape};
\node (CL1)[right of = reshape,xshift=-0.5cm,yshift=0.4cm]  
    {$C_L$};
\node (decN1a) [process,minimum width=1.6cm,right of = reshape,xshift=1.3cm,fill=red!30,rotate=90] {ConvT (T,F)};
\node (CL11)[right of = decN1a,xshift=-0.24cm,yshift=0.4cm]  
    {$C_{L-1}$};
\node (C21)[right of = decN1a,xshift=0.8cm,yshift=0.4cm]  
    {$C_2$};
\node (decN1b) [process,minimum width=2.1cm,right of = decN1a,xshift=1.5cm,fill=red!30,rotate=90] {ConvT (T,F)};
\node (C11)[right of = decN1b,xshift=-0.3cm,yshift=0.4cm]  
    {$C_1$};
\node (decN1c) [process,minimum width=2.6cm,right of = decN1b,xshift=1cm,fill=red!30,rotate=90] {ConvT (T,F)};
\node (Cout)[right of = decN1c,xshift=-0.3cm,yshift=0.4cm]  
    {$C_{out}$};
\node (decSa) [process,minimum width=1.6cm,above of = decN1a,yshift=2cm,fill=green!30,rotate=90] {ConvT (T,F)};
\node (decSb) [process,minimum width=2.1cm,right of = decSa,xshift=1.5cm,fill=green!30,rotate=90] {ConvT (T,F)};
\node (decSc) [process,minimum width=2.6cm,right of = decSb,xshift=1cm,fill=green!30,rotate=90] {ConvT (T,F)};
\node (decN2a) [process,minimum width=1.6cm,below of = decN1a,yshift=-2.4cm,fill=red!30,rotate=90] {ConvT (T,F)};
\node (decN2b) [process,minimum width=2.1cm,right of = decN2a,xshift=1.5cm,fill=red!30,rotate=90] {ConvT (T,F)};
\node (decN2c) [process,minimum width=2.6cm,right of = decN2b,xshift=1cm,fill=red!30,rotate=90] {ConvT (T,F)};
\node (dot0) [c,right of = reshape, xshift = 0.5 cm, fill = black,minimum size=0.5em]{};
\node (dot1) [c,right of = enc1, xshift = 0 cm, fill = black,minimum size=0.5em]{};
\node (dot2) [c,right of = enc2, xshift = -0.15 cm, fill = black,minimum size=0.5em]{};
\node (dot3) [c,right of = enc3, xshift = 0 cm, fill = black,minimum size=0.5em]{};
\node (add1) [c,minimum size=1em,right of = reshape,xshift=-0.12cm] {$+$};
\node (add2) [c,minimum size=1em,right of = decN1a,xshift=0.cm] {$+$};
\node (add3) [c,minimum size=1em,right of = decN1b,xshift=0.cm] {$+$};
\node (addN2a) [c,minimum size=1em,right of = decN2a,xshift=0.cm] {$+$};
\node (addN2b) [c,minimum size=1em,right of = decN2b,xshift=0.cm] {$+$};
\node (addSa) [c,minimum size=1em,right of = decSa,xshift=0.cm] {$+$};
\node (addSb) [c,minimum size=1em,right of = decSb,xshift=0.cm] {$+$};
\node (S)[right of = decSc,xshift=0cm] {$\hat{S}$};
\node (N1)[right of = decN1c,xshift=0cm] {$N_1$};
\node (N2)[right of = decN2c,xshift=0cm] {$N_2$};
\draw [arrow] (Y) -- (enc1);
\draw [arrow] (enc1) -- (enc2);
\draw [dotted,arrow] (enc2) -- (enc3);
\draw [arrow] (enc3) -- (flatten1);
\draw [arrow] (reshape) -- (decN1a);
\draw [dotted,arrow] (decN1a) -- (decN1b);
\draw [arrow] (decN1b) -- (decN1c);
\draw [dotted,arrow] (decSa) -- (decSb);
\draw [arrow] (decSb) -- (decSc);
\draw [dotted,arrow] (decN2a) -- (decN2b);
\draw [arrow] (decN2b) -- (decN2c);
\draw [arrow] (dot0) |- (decSa);
\draw [arrow] (dot0) |- (decN2a);
\draw [arrow] (decSc) -- (S);
\draw [arrow] (decN1c) -- (N1);
\draw [arrow] (decN2c) -- (N2);
\draw [dashed,arrow] ($(enc1) + (0.3,1.4)$) -- ($(enc2) + (-0.4,1.1)$);
\draw (dot3) -- ($(dot3)+(0,-1.5)$); 
\draw [arrow] ($(add1)+(0,-1.5)$) --(add1); 
\draw ($(add1)+(0,-1.5)$) -- ($(dot3)+(0,-1.5)$); 
\draw (dot2) -- ($(dot2)+(0,-1.7)$); 
\draw [arrow] ($(add2)+(0,-1.7)$) --(add2); 
\draw ($(add2)+(0,-1.7)$) -- ($(dot2)+(0,-1.7)$);
\draw (dot1) -- ($(dot1)+(0,-1.9)$); 
\draw [arrow] ($(add3)+(0,-1.9)$) --(add3); 
\draw ($(add3)+(0,-1.9)$) -- ($(dot1)+(0,-1.9)$);
\draw (dot2) -- ($(dot2)+(0,-4.5)$); 
\draw ($(dot2)+(0,-4.5)$) -| (addN2a);
\draw (dot1) -- ($(dot1)+(0,-5)$); 
\draw ($(dot1)+(0,-5)$) -| (addN2b);
\draw (dot1) -- ($(dot1)+(0,5)$); 
\draw ($(dot1)+(0,5)$) -| (addSb);
\draw (dot2) -- ($(dot2)+(0,4.5)$); 
\draw ($(dot2)+(0,4.5)$) -| (addSa);
\draw (CL) -- (enc3);
\end{tikzpicture}
\end{adjustbox}
\caption{System encoder, bottleneck and decoder . There are $L$ convolutional encoder and $L$ deconvolutional decoders layers. The encoder and decoder layers are symmetric to each others. The bottleneck includes $J$ GRU layers, flatten and reshape operations }
\label{fig:detailedNetwork}
\end{figure}
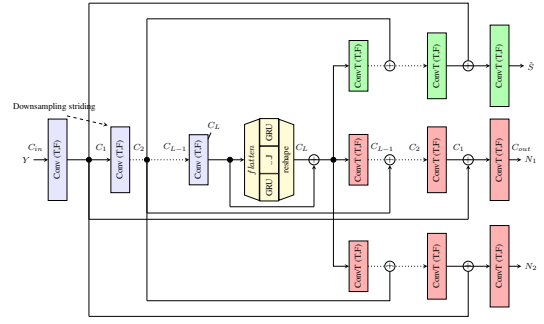


\subsubsection{Unsupervised architecture}
The network architecture for unsupervised MixIT training is identical to the supervised case, except that we add two additional decoder branches for $N_1$ and $N_2$ as shown in Fig.~\ref{fig:detailedNetwork}. The two additional decoders for the noise outputs are required for training only, adding no additional inference complexity. We added a pre-trained wav2vec2 model \cite{baevski2020wav2vec} to extract ASR embeddings.
The input features are extracted from the mixture $Y$, where to the noisy speech $X$ the noise $N$ is added as shown in Fig.~\ref{fig:unsup}. The network then predicts three filters to be applied to the input to obtain the spectra $\hat{S},N_1,N_2$.

\subsection{Semi-supervised speech enhancement}
An overall benefit could be achieved by joining supervised and unsupervised training methods into a semi-supervised approach. This means, we train on both clean and noisy speech datasets, using the supervised and unsupervised losses in parallel. To train the semi-supervised network, we feed a batch of utterances from clean dataset to calculate the supervised loss, and feed a batch of utterances from noisy dataset to calculate the unsupervised loss. The semi-supervised training loss is the sum of supervised loss from \eqref{eq:eqsup} and unsupervised loss from \eqref{eq:equnsup}, and is given by
\begin{align}
    \label{eq:semisup}
    \mathcal{L}_{semi} = \mathcal{L}_{supervised} + \mathcal{L}_{unsupervised}.
\end{align}
The network weights are updated on the joint semi-supervised loss.

\section{Experimental Setup}
\subsection{Dataset}
\begin{figure}[t]
\centering  
\includegraphics[width=0.85\columnwidth,clip,trim=10 0 10 10]{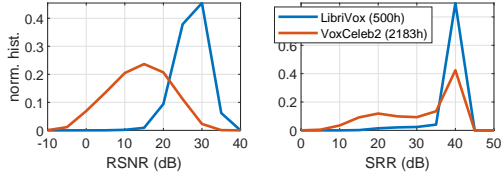}  
\caption{Distributions of RSNR and SRR in the LibriVox and VoxCeleb datasets}
\label{fig:SNR}
\end{figure}
We used the LibriVox \footnote[1]{https://librivox.org/} with 500 hours as a clean training dataset for the supervised baseline models. We used the VoxCeleb2 \cite{chung2018voxceleb2} dataset as noisy speech training data for unsupervised training. The distributions of reverberant signal-to-noise ratio (RSNR)  and signal-to-reverberation ratio (SRR) of the utterances in these two datasets are measured using a blind predictor based on \cite{braun2021on} and visualized in Figure \ref{fig:SNR}. We can see that VoxCeleb2 has a lot of low SNR content and also reverberant speech, compared to LibriVox.
The noise added to the speech in training stage is drawn from the of DNS challenge noise data\footnote[2]{https://github.com/microsoft/DNS-Challenge/tree/icassp2021-final} (180 hours). The development set contains 300 synthetic noisy utterances by mixing clean speechs from DAPS \footnote[3]{https://ccrma.stanford.edu/ gautham/Site/daps.html} corpus with noises and impulse responses from the QUT \footnote[4]{https://research.qut.edu.au/saivt/databases/qut-noise} database . The speech enhancement test set are 600 utterances both verbal and non-verbal audio recordings from the 3rd DNS challenge \cite{reddy21_interspeech} test set. 

For supervised training, non-reverberant recordings from LibriVox were augmented with impulse responses (IRs) from the DNS challenge (115k IRs), and from QUT in the training set and development set respectively. We created either fully reverberant or non-reverberant speech training targets as described in \cite{braun2021towards} to create different supervised baselines. 

The speech recognition test set is the DNS challenge test set. We employed the public Azure speech software development kit transcription service to get the ASR performance. 
The WER is computed on three test sets: i) the 600 clips from the 3rd DNS challenge \cite{reddy21_interspeech}, ii) 200 clips of high SNR speech to check degradations when noise suppression is not necessary, and iii) an internal test set 18~h actual meeting recordings.


\subsection{Evaluation}
Our goal is to improve speech enhancement without degrading ASR performance. Thus, we evaluate our system with both speech enhancement and ASR metrics. Speech enhancement systems can be evaluated on objective or subjective human ratings. As the human rating is expensive and slow, neural networks are used to predict the human mean opinion score (MOS) \cite{reddy2021dnsmos,gamper2019intrusive}.  
The ITU P.835 standard rates the MOS of 3 components, signal distortion (SIG), background noise suppression (BAK) and overall quality (OVL) with scores ranging from 1 to 5, the higher the better. We use a neural network based on  \cite{reddy2021dnsmos} to predict the non-intrusive scores \emph{nSIG}, \emph{nBAK}, \emph{nOVL}, where "n" stands for non-intrusive. The speech recognition performance is evaluated using the standard word error rate (WER) metric.
\begin{table*}[tb]
  \footnotesize
  \caption{Speech enhancement and speech recognition scores on the DNS challenge test set} \label{tbl:result1}
  \centering
  \begin{tabular}{llllllp{11mm}p{12mm}p{11mm}}
    \toprule
    \textbf{Exp} & \textbf{Method} &   \textbf{Dataset}   & \textbf{nSIG}& \textbf{nBAK}& \textbf{nOVL}& \textbf{WER \newline low SNR}& \textbf{WER \newline high SNR} & \textbf{WER \newline meeting} \\    
    \midrule
    \rule{0pt}{2ex}
    & No enhancement  & None &  3.87	& 3.05 & 3.11 &	27.95	& 5.69 & 16.52 \\
    Exp 1  & Supervised (lower bound)  & Noisy  & 3.75 &	3.97	&3.31
 & 35.51 &	6.81 &	20.03
 \\
    Exp 2 & Supervised (upperbound) & Clean + dereverb  &3.76&	4.27	&3.52 &	32.59	&6.15 &	19.21 \\
    Exp 3 & Supervised (baseline)  & Clean + Reverb & 3.75 &	4.12 &	3.39 	& 32.37 &	5.97 &	19.29 \\
    Exp 4 & MixIT & Noisy &  3.80 &	3.28  &	3.16  &	31.61  &	5.74  &	17.16 \\
    Exp 5 & MixIT + ASR embedding (Emb) & Noisy & 3.67 &	3.93 &	3.25	& 31.09 &	6.52 & 19.65
 \\
    Exp 6 & MixIT + disentanglement  (Dis)& Noisy & 3.70	& 3.97 &	3.27 &	32.51	&6.74	& 20.51
 \\
    Exp 7 & MixIT + Emb + Dis & Noisy &  3.69	&4.00 &	3.29	&33.58	&6.44 &	21.13 \\
Exp 8 &	Semisupervised with Emb &	Clean + noisy&	3.76	&4.23 &	3.49	&30.94	&6.11	&18.77
 \\
Exp 9 &	Semisupervised with Emb + Dis &	Clean + noisy &	3.76 &	4.22	& 3.49 &	32.22	& 5.89 &	18.86
 \\
    \bottomrule
  \end{tabular}
\end{table*}

\subsection{Training parameters}
The encoder has 4 layers with 16-32-64-128 channels. Each decoder has 4 layers with 128-64-32-16 channels. The pretrained Wav2vec 2.0 is Wav2Vec 2.0 large (LV-60) \cite{baevski2020wav2vec} with self-training. The training utterances have the length of 10 seconds. 

The network was trained with an AdamW optimizer with a learning rate of $1\cdot 10^{-3}$ for the supervised loss, $5 \cdot 10^{-4}$ for the unsupervised loss, and weight decay $2\cdot10^{-5}$. The learning rate is reduced by half if there is no improvement in the development set performance after 200 epochs. The epoch size is 5120. The loss weights are $\lambda=0.3$, $\alpha_\text{e}=0.004$, $\alpha_\text{d}=0.0005$. 

We measured the performance of all the models on the development set every $E$ epochs and choose the best model based on a weighted metric following \cite{braun2021towards}. The metric is a weighted combination of cepstral distance (CD), perceptual evaluation of speech quality (PESQ) \cite{rix2001perceptual} and scale-invariant signal-to-
distortion ratio: 
\begin{align}
    M = PESQ + 0.2siSDR -CD
     \label{eq:eq2}
\end{align}

\subsection{Experiments}
To have fair comparisons and to gauge upper and lower bounds, we first set up three supervised baseline experiments. In Exp.~1 (lower bound), the speech signal $S$ is drawn from the noisy speech dataset (VoxCeleb2) and noise $N$ is taken randomly from the training noise dataset. This experiment acts as lower bound because we expect the unsupervised loss could improve the trivial supervised training on noisy speech. In Exp.~2 (upper bound), the speech signal $S$ is from the clean speech dataset LibriVox and is augmented with impulse responses (RIRs). The target speech is created by using windowed RIRs to obtain non-reverberant speech, i.\,e,\ the network also learns dereverberation. Exp.~3 is identical to Exp.~2, except that the speech training targets are kept fully reverberant. Exp.~3 acts as a fair baseline, as the MixIT loss has no incentive to learn dereverberation.  


In Exp.~4, we implemented the original MixIT training \cite{NEURIPS2020_28538c39} in a fully unsupervised fashion from noisy speech using the loss \eqref{eq:mixit}. 
Exp.~7 uses the proposed loss \eqref{eq:equnsup}, adding both embedding and disentanglement losses to the MixIT loss. In addition, we conducted an ablation study to investigate the contribution of disentanglement or ASR embedding losses. We removed the disentanglement or ASR embedding losses in Exp.~5, 6 by setting either $\alpha_\text{d}$ or $\alpha_\text{e}$ in \eqref{eq:equnsup} to zero. 

We also did two semi-supervised experiments (Exp.~8, 9), combining supervised and unsupervised loss by training on \eqref{eq:semisup}. In Exp.~8, we train on MixIT with ASR embedding loss ($\alpha_\text{d}=0$), while Exp.~9, is the full semi-supervised combination of MixIT, ASR embedding and disentanglement loss. 
  

\section{Results}


The experiment results are shown in Table \ref{tbl:result1}. The nSIG, nBAK, nOVL metrics are computed on the 3rd DNS challenge test set, while the WER is shown the 3 test sets DNS challenge (\emph{low SNR}), the high SNR test set (\emph{high SNR}), and the meeting recordings (\emph{meeting}).
The first line presents the metrics of the noisy speech utterances in the test sets without applying any speech enhancement methods. The noisy speech has the lowest nBAK score (3.05) compared to the enhanced signals in other experiments because it contains a lot of noise. It also has the lowest nOVL score (3.11). However, it has the lowest WERs because applying speech enhancement systems to remove the noise led to signal distortion. We measure the WER on an ASR which is not jointly trained with a speech enhancement system. 

The results of supervised experiments are described from row 2 to 4. Exp.~2 has the highest speech enhancement overall (nOVL) score at 3.52, because it is trained on clean, non-reverberant speech targets, i.\,e.\ reducing both noise and reverberation. As we can see, a  model trained with strong supervision on a large dataset is still hard to be outperformed. The nOVL score decreases to 3.39 when the training speech target has reverberation in Exp.~3, i.\,e.\ where the network is trained to remove noise only, but no reverberation. In Exp.~1, the nOVL drops to 3.31 when using noisy speech as target.  

The unsupervised training experiments results are reported in Exp.~4 to 7. In Exp.~4, the original MixIT loss shows a significantly worse enhancement performance with its nOVL (3.16) being slightly above the noisy speech without enhancement (row 1) nOVL (3.11) and nBAK of only 3.28. MixIT also has WER for the meeting and high SNR datasets close to the noisy speech row 1 (17.16\% vs 16.52\% and 5.74\% vs 5.69\%). MixIT has higher low SNR WER than the noisy speech in row 1 (31.61\% vs 27.95\%). 

The proposed unsupervised loss \eqref{eq:equnsup} in Exp.~7 has a similar nOVL and a better WER compared to the supervised training loss on Exp.~1. 
It has better speech recognition performance on low SNR, high SNR test cases with a lower error rate at 33.58\%, 6.44\% compared to 35.51\%, 6.81\% of Exp.~1, respectively.  

The ablation study (Exp.~5, 6) illustrates how the performance changed when we dropped the disentanglement or the ASR embedding. As we can see, when the disentanglement or ASR embedding was dropped, the speech enhancement performance degrades (nOVL drops from 3.29 to 3.25 and 3.27 respectively). Thus, the disentanglement and ASR embedding loss terms show to be helpful constraints on the system. 

However, the fully unsupervised training could not exceed the baseline (Exp.~3), a strong supervised training with reverberation. An interesting question is if the semi-supervised training can exceed the upper supervised bound (Exp.~2). Exp.~8 has an nOVL comparable to the supervised training which uses clean speech as target (Exp.~2), and better WERs. More specifically, Exp.~8 has a WER at 30.94 \%, 6.11 \% , 18.77 \%  compared to 32.59 \%, 6.15\%, 19.21\% of Exp.~2 on the low SNR, high SNR and meeting datasets, respectively. While the semi-supervised training could not outperform the supervised upper bound in terms of MOS, likely due to the that fact the unsupervised training samples will not push for dereverberation,  it exceeds the reverberant baseline (Exp.~3) and shows less WER degradation than the supervised upper bound.

\section{Conclusion}
In this work, we improved the MixIT loss, which is an unsupervised training scheme for speech enhancement using only noisy speech. While the original MixIT loss was shown to underperform in terms of noise suppression, we proposed two additional terms, an ASR embedding loss and a disentanglement loss to improve ASR performance and enforce the speech from noise separation.
With the additions, our improved unsupervised system could outperform a supervised baseline trained on noisy speech, but not a supervised baseline trained on clean reverberant speech.
We therefore proposed a semi-supervised training scheme, jointly training with a supervised loss on clean speech, and with MixIT on noisy speech. This method could outperform the supervised baseline trained on clean reverberant speech. However it could not outperform the strongest upper bound supervised baseline trained on clean non-reverberant speech in terms of MOS, but in terms of ASR performance.

In future work, the approach can be extended by adding different embedding related loss to the proposed loss function. Another interesting topic is maintaining important speech cues for ASR \cite{trinh2022importantaug, trinh18_interspeech,DBLP:conf/interspeech/TrinhM20} when doing speech enhancement to minimize the speech enhancement vs.\ speech recognition trade-off. 


\vfill\pagebreak
\balance


\bibliographystyle{IEEEbib}
\bibliography{strings,refs}

\begin{thebibliography}{10}

\bibitem{weninger2015speech}
Felix Weninger, Hakan Erdogan, Shinji Watanabe, Emmanuel Vincent, Jonathan
  Le~Roux, John~R Hershey, and Bj{\"o}rn Schuller,
\newblock ``Speech enhancement with lstm recurrent neural networks and its
  application to noise-robust asr,''
\newblock in {\em International conference on latent variable analysis and
  signal separation}. Springer, 2015, pp. 91--99.

\bibitem{luo2019conv}
Yi~Luo and Nima Mesgarani,
\newblock ``Conv-tasnet: Surpassing ideal time--frequency magnitude masking for
  speech separation,''
\newblock {\em IEEE/ACM transactions on audio, speech, and language
  processing}, vol. 27, no. 8, pp. 1256--1266, 2019.

\bibitem{reddy21_interspeech}
Chandan~K.A. Reddy, Harishchandra Dubey, Kazuhito Koishida, Arun Nair, Vishak
  Gopal, Ross Cutler, Sebastian Braun, Hannes Gamper, Robert Aichner, and
  Sriram Srinivasan,
\newblock ``{INTERSPEECH 2021 Deep Noise Suppression Challenge},''
\newblock in {\em Proc. Interspeech 2021}, 2021, pp. 2796--2800.

\bibitem{li2021icassp}
Andong Li, Wenzhe Liu, Xiaoxue Luo, Chengshi Zheng, and Xiaodong Li,
\newblock ``Icassp 2021 deep noise suppression challenge: Decoupling magnitude
  and phase optimization with a two-stage deep network,''
\newblock in {\em ICASSP 2021-2021 IEEE International Conference on Acoustics,
  Speech and Signal Processing (ICASSP)}. IEEE, 2021, pp. 6628--6632.

\bibitem{wang2018supervised}
DeLiang Wang and Jitong Chen,
\newblock ``Supervised speech separation based on deep learning: An overview,''
\newblock {\em IEEE/ACM Transactions on Audio, Speech, and Language
  Processing}, vol. 26, no. 10, pp. 1702--1726, 2018.

\bibitem{tan2019learning}
Ke~Tan and DeLiang Wang,
\newblock ``Learning complex spectral mapping with gated convolutional
  recurrent networks for monaural speech enhancement,''
\newblock {\em IEEE/ACM Transactions on Audio, Speech, and Language
  Processing}, vol. 28, pp. 380--390, 2019.

\bibitem{NEURIPS2020_28538c39}
Scott Wisdom, Efthymios Tzinis, Hakan Erdogan, Ron Weiss, Kevin Wilson, and
  John Hershey,
\newblock ``Unsupervised sound separation using mixture invariant training,''
\newblock in {\em Advances in Neural Information Processing Systems},
  H.~Larochelle, M.~Ranzato, R.~Hadsell, M.~F. Balcan, and H.~Lin, Eds. 2020,
  vol.~33, pp. 3846--3857, Curran Associates, Inc.

\bibitem{saito2021training}
Koichi Saito, Stefan Uhlich, Giorgio Fabbro, and Yuki Mitsufuji,
\newblock ``Training speech enhancement systems with noisy speech datasets,''
\newblock {\em arXiv preprint arXiv:2105.12315}, 2021.

\bibitem{wang2020self}
Yu-Che Wang, Shrikant Venkataramani, and Paris Smaragdis,
\newblock ``Self-supervised learning for speech enhancement,''
\newblock {\em arXiv preprint arXiv:2006.10388}, 2020.

\bibitem{fujimura2021noisy}
Takuya Fujimura, Yuma Koizumi, Kohei Yatabe, and Ryoichi Miyazaki,
\newblock ``Noisy-target training: A training strategy for dnn-based speech
  enhancement without clean speech,''
\newblock {\em arXiv preprint arXiv:2101.08625}, 2021.

\bibitem{braun2021consolidated}
Sebastian Braun and Ivan Tashev,
\newblock ``A consolidated view of loss functions for supervised deep
  learning-based speech enhancement,''
\newblock in {\em 2021 44th International Conference on Telecommunications and
  Signal Processing (TSP)}. IEEE, 2021, pp. 72--76.

\bibitem{braun2021towards}
Sebastian Braun, Hannes Gamper, Chandan~KA Reddy, and Ivan Tashev,
\newblock ``Towards efficient models for real-time deep noise suppression,''
\newblock in {\em ICASSP 2021-2021 IEEE International Conference on Acoustics,
  Speech and Signal Processing (ICASSP)}. IEEE, 2021, pp. 656--660.

\bibitem{cho2014properties}
Kyunghyun Cho, Bart van Merri{\"e}nboer, Dzmitry Bahdanau, and Yoshua Bengio,
\newblock ``On the properties of neural machine translation: Encoder--decoder
  approaches,''
\newblock in {\em Proceedings of SSST-8, Eighth Workshop on Syntax, Semantics
  and Structure in Statistical Translation}, 2014, pp. 103--111.

\bibitem{baevski2020wav2vec}
Alexei Baevski, Yuhao Zhou, Abdelrahman Mohamed, and Michael Auli,
\newblock ``wav2vec 2.0: A framework for self-supervised learning of speech
  representations,''
\newblock {\em Advances in Neural Information Processing Systems}, vol. 33,
  2020.

\bibitem{chung2018voxceleb2}
Joon~Son Chung, Arsha Nagrani, and Andrew Zisserman,
\newblock ``Voxceleb2: Deep speaker recognition,''
\newblock {\em Proc. Interspeech 2018}, pp. 1086--1090, 2018.

\bibitem{braun2021on}
Sebastian Braun and Ivan Tashev,
\newblock ``On training targets for noise-robust voice activity detection,''
\newblock in {\em European Signal Processing Conference (EUSIPCO)}, August
  2021.

\bibitem{reddy2021dnsmos}
Chandan~KA Reddy, Vishak Gopal, and Ross Cutler,
\newblock ``Dnsmos: A non-intrusive perceptual objective speech quality metric
  to evaluate noise suppressors,''
\newblock in {\em ICASSP 2021-2021 IEEE International Conference on Acoustics,
  Speech and Signal Processing (ICASSP)}. IEEE, 2021, pp. 6493--6497.

\bibitem{gamper2019intrusive}
Hannes Gamper, Chandan~KA Reddy, Ross Cutler, Ivan~J Tashev, and Johannes
  Gehrke,
\newblock ``Intrusive and non-intrusive perceptual speech quality assessment
  using a convolutional neural network,''
\newblock in {\em 2019 IEEE Workshop on Applications of Signal Processing to
  Audio and Acoustics (WASPAA)}. IEEE, 2019, pp. 85--89.

\bibitem{rix2001perceptual}
Antony~W Rix, John~G Beerends, Michael~P Hollier, and Andries~P Hekstra,
\newblock ``Perceptual evaluation of speech quality (pesq)-a new method for
  speech quality assessment of telephone networks and codecs,''
\newblock in {\em 2001 IEEE international conference on acoustics, speech, and
  signal processing. Proceedings (Cat. No. 01CH37221)}. IEEE, 2001, vol.~2, pp.
  749--752.

\bibitem{trinh2022importantaug}
Viet~Anh Trinh, Hassan~Salami Kavaki, and Michael~I Mandel,
\newblock ``Importantaug: a data augmentation agent for speech,''
\newblock {\em ICASSP 2022 IEEE International Conference on Acoustics, Speech
  and Signal Processing (ICASSP)}, 2022.

\bibitem{trinh18_interspeech}
Viet~Anh Trinh, Brian McFee, and Michael~I Mandel,
\newblock ``{Bubble Cooperative Networks for Identifying Important Speech
  Cues},''
\newblock in {\em Proc. Interspeech 2018}, 2018, pp. 1616--1620.

\bibitem{DBLP:conf/interspeech/TrinhM20}
Viet~Anh Trinh and Michael~I. Mandel,
\newblock ``Large scale evaluation of importance maps in automatic speech
  recognition,''
\newblock in {\em Interspeech 2020, 21st Annual Conference of the International
  Speech Communication Association}. 2020, pp. 1166--1170, {ISCA}.

\end{thebibliography}

\end{document}